\documentstyle[12pt,aasms]{article}
\lefthead{Chakrabarti}
\righthead{Line Profiles in M87}

\begin{document}

\title{Velocity Profile of the Ionized Disk and the Mass of the
Black Hole in M87}

\author{Sandip K.\ Chakrabarti}
\affil{Max-Planck-Institut f\"ur Astrophysik, Garching bei M\"unchen, \\
Karl Schwarzschild Str. 2, D-85740 Germany}
\centerline {and}
\affil{Theoretical Astrophysics, Tata Institute of Fundamental
Research{\altaffilmark{1}},
\\ Homi Bhabha Road, Bombay 400005, India \ \
 e-mail: chakraba@tifrvax.tifr.res.in}

\altaffiltext{1}{Permanent Address}

\begin{abstract}

We present a theoretical model for the ionized disk in M87
which includes spiral shock waves. The line emission profiles
computed from this model at various positions on the disk are found to be
in agreement with the recent Hubble Space Telescope results.
Based on this model, we find that the ionized disk comprises two-armed giant
spiral shock waves which extend from around $0.1$ arc sec from the center
to at least $1$ arc second or more. Our model requires that the mass of the
black hole be $(4 \pm 0.2) \times 10^9 M_\odot$ and the inclination angle
to be $(42\pm 2)^{\deg}$. We predict the nature of the line profiles at many
other locations of the disk which could be verified in future observations.
\end{abstract}



\section{Introduction}

Using the Faint Object Spectrograph (FOS) on Hubble Space Telescope (HST),
Harms et al. (1994, hereafter H94) have recently reported the spectroscopy of
central region of the elliptical galaxy M87. Ford et al. (1994, hereafter F94)
using Wide Field Planetary Camera-2 have imaged the disk in $H_\alpha$+[NII]
and find an ionized disk with spiral structures of two (or more) arms.
{}From the kinematical arguments, based on the Doppler shifts
of several lines emitted from the disk, and assuming
a Keplerian motion of the emitting gas, they conclude that the  mass of the
disk plus the nucleus: $M_c(R <18$pc$)= (2.4 \pm 0.7) \times
10^9 M_\odot$ and the inclination angle of the disk with the line of
sight is $i= (42 \pm 5)^{\deg}$.

In the present paper, we provide a complete description of the velocity field
of the ionized disk and compute shapes of typical line profiles expected
from various regions of the disk. Our analysis is based on the solution
of a non-axisymmetric disk model which includes two armed spiral density waves
possibly observed in the HST data and therefore our disk model is intrinsically
non-Keplerian. We find a very good agreement between the theoretical
and observed line profiles as regards to the Doppler shifts, line widths and
the intensity ratios. We generally confirm the conclusions
of F94 and H94 regarding the mass of the black hole and the inclination of the
disk although we believe that the mass of the central black hole
is somewhat higher --- close to $4 \times 10^9 M_\odot$. We interprete
this to be due to the underestimation of the mass
in previous reports which assume Keplerian motions of the emitting gas.

Based on the observed $H_\alpha$ flux and the electron density,
F94 computes the mass of the ionized disk to be at most $10^4M_\odot$
which is small compared with the central black hole. Assuming
that the neutral disk mass to be of same order, it is unlikely that
the observed spiral feature is due to the self-gravity of the disk.
In a binary system with a thin accretion disk, the
companion can induce two armed spiral shocks in the disk (e.g.,
Matsuda et al. 1987, Spruit 1987, Chakrabarti \& Matsuda, 1992).
In the case of active galaxies, a passing companion or even a globular
cluster which is more massive than the disk can induce the same effect, but
the spiral structure need not be stationary (Chakrabarti \& Wiita 1993).
The lifetime of the spiral structures may be of the order of a few Keplerian
periods after which they break up and disappear at higher radius. It is
possible
that one such broken piece of the spiral shock structure may be
visible in Fig. 3 of the M87 disk (F94). This figure
indicates the existence of a third spiral arm which is presently
no longer attached to the nucleus and is about half an arc second away.
Because of this, we concentrate on disk solutions which include
only two spirals. We have tested our solution
with three arms as well but do not find satisfactory results
to  match with the observations.

The spiral shock wave solution is described in detail
in the literature (Chakrabarti, 1990ab) and therefore it will not be presented
here. The general procedure of obtaining the intensity of line emissions
from such a disk is also presented elsewhere (Chakrabarti \& Wiita, 1994)
where it was shown that the time variability of the line emissions
from the broad line radio galaxies such as ARP 102B and 3C390.3
can be understood well using non-Keplerian disk with spiral shock
waves. In Section 2, we present the model assumptions.
We also briefly discuss the analytical model and the procedure followed
to obtain the disk line emission. In Section 3, we present
the assumptions which went into fitting the observed line profiles in M87.
In Section 4, we produce the parameters which yield the best fit and compare
line profiles obtained from our disk model with those presented in H94.
We also predict the nature of the profiles in regions not yet
observed by HST. Finally, in Section 5, we present our conclusions.

\section {Model Equations}

Since we shall be discussing physical processes far away from the
black hole, a Newtonian theory of the spiral shock wave in a
quasi-three dimensional accretion disk will suffice.
A general theory of such  spiral shocks in the context of
non-relativistic accretion disks is presented in Chakrabarti
(1990a, hereafter Paper 1) and that including relativistic corrections
is presented in Chakrabarti \& Wiita (1994). In these works,
for simplicity, the spiral structure is assumed to be self-similar, i.e.,
the angle subtended by the arms with the radial vector is assumed
to be constant with radial coordinate.

In a self-similar disk, the velocity components vary as,
\begin{equation}
u^{\prime}=x^{-1/2} q_1 (\Psi), \ \ \
v^{\prime}=x^{-1/2} q_2 (\Psi), \ \ {\rm and} \ \
a^{\prime}=x^{-1/2} q_3 (\Psi),
\end{equation}
where, $u^{\prime}$, $v^{\prime}$, and $a^{\prime}$ are the radial, azimuthal
and sound velocities, respectively,  $\Psi=\phi^{\prime}+B \
$log$(x) = constant$
defines the spiral coordinate, $x$ is the radial co-ordinate $r^{\prime}$
in units of $GM/c^2$, $M$ being the mass of the central black hole,
and $B = $tan$\, \beta$, with $\beta$ as the pitch angle of the
spirals. The functional behaviors of $q_i (\Psi)$  are self-consistently
determined from the two-and-a-half dimensional Euler equations:
all the terms on the equatorial $(r^{\prime}, \ \phi^{\prime} )$ plane are
kept, while the advective terms are dropped in the vertical direction
and a hydrostatic equilibrium is assumed in that direction.
The assumption of hydrostatic equilibrium may break down in the post
shock region as the flow becomes hotter and expands in the transverse
direction. However the gravitational influence of the central object
is expected to extend about a distance of $300$pc and unless the
disk violently oscillates, the vertical velocity component is not expected to
be
high compared to the radial or azimuthal component. Therefore, the assumption
of hydrostatic equilibrium is a good one to make under the circumstances.
We assume the flow to obey an equation of state: $P=K(x,\ \Psi)\rho^\gamma$,
where, $P$ and $\rho$ are pressure and density respectively, and $\gamma=
1+1/n$ is a constant, with $n$ as the polytropic index.
For an assumed number of shocks, given the pitch angle $\beta$, and any one of
the velocity components on the sonic surface (we supply $q_{2c}$), the entire
set of equations can be solved for all the kinematic quantities $q_i(\Psi)$
as well as the polytropic index $n$ of the disk. For our purpose in this paper,
knowledge of these quantities are sufficient to obtain the velocity fields
and line profiles. In a self-similar disk in
hydrostatic equilibrium in the vertical direction, pressure, density and
the entropy function $K$ are found to vary as (Paper 1):
\begin{equation}
P=x^{-5/2} q_p (\Psi), \ \ \
\rho=x^{-3/2} q_\rho (\Psi), \ \ {\rm and} \ \
K=x^{\frac{3\gamma}{2}-\frac{5}{2}} \frac{q_p}{q_\rho^\gamma} \\
\end{equation}
For a complete knowledge of the thermodynamic quatities, one has to
supply another boundary condition, say $q_\rho$ at a point. This, together
with Eqn. (2) and the definition of the sound speed $a=(\gamma P/\rho)^{1/2}$,
one can obtain all the necessary quantities. One important point to note
is that since entropy must increase as the matter accretes (as $x$ decreases),
one must consider only those solutions which yield $\gamma < 5/3$, or $n >3/2$.
A few examples of the spiral shock solutions are given
in Paper 1. For the sake of convenience $q_3$ as defined here is square
root of $q_3$ of Paper 1. For comparison: in a strictly Keplerian flow,
$q_2=1$ and $q_1=0$ identically everywhere. The sound speed in
a Keplerian disk varies as $a \propto x^{-3/8}$, so $q_3$ cannot be compared
directly with the Keplerian value.

The velocity components in above equations are transformed to the observers
frame and relativistic corrections are made
following Chen, Helpern \& Fillipenko (1989, hereafter CHF) and
Chakrabarti \& Wiita (1994). We assume a power law emissivity,
\begin{equation}
\epsilon (x) \propto x^{-q},
\end{equation}
over the boundaries ($x_{in}$ and $x_{out}$) of the line emitting region of
the disk. The index $q$ is expected to be within the range
$1 \leq q \leq 3$ (e.g., Tylenda 1981; Smak 1981; CHF).
The intensity of line emission at a given frequency is then obtained,
following  the approach of CHF, to be
\begin{equation}
I_\nu \propto \int_0^\infty \epsilon(x) \frac{D^3}{v_{th}} {\rm exp}
[ -\frac{1}{2}(\frac{\nu/D - 1}{v_{th}})^2] d\Omega,
\end{equation}
where $v_{th}$ characterizes the line broadening due to thermal motions
of the emitting particles which we choose to be constant
for simplicity. Here, $D=\nu/\nu_e$, is the Doppler factor with $\nu_e$
as the emitted frequency and $\nu$ as the observed frequency.
The solid angle $d \Omega$ is given by $b \ db \ d\phi /d^2$, where
$b$ is the impact parameter of the photon (Chakrabarti \& Wiita 1994; CHF)
and $d$ is the distance to the AGN. Instead of using constant $v_{th}$,
one could also choose the thermal velocity,
$v_{th} \propto (u^2+v^2)^{1/2} h(x)/x = Q_{th}
q_3 (q_1^2+q_2^2)^{1/2} x^{-1/2}$, where $h(x)=x q_3$ is the local
half-thickness of the disk and $Q_{th}$ is a proportionality
constant. The resulting line profiles with a suitable $Q_{th}$
appear to be very similar as those with a constant $v_{th}$.

\section{Fitting Procedure of the HST Results}

For a given system, the general procedure of fitting the line profiles with
the theoretical result would be to supply the following quantities: $B$
and $q_{2c}$ to obtain the velocity profiles in the disk, and the inclination
angle $i$, the inner and outer edges of the emitting regions, $x_{in}$
and $x_{out}$ respectively, the emissivity law $q$, thermal broadening
velocity $v_{th}$ and the mass of the black hole $M$
to obtain the line intensities. Among the
seven positions of the disk pointed at by HST (see Fig. 1b
of H94), we shall be interested in results
obtained from the regions marked as positions 1, 2, 4, 5 and 6
since the emission line profiles are prominent in these regions.
Figure 1 schematically shows these regions as
dotted circles. Each of the circular apertures
is of diameter $0.26$ arc sec which corresponds to an absolute
length of $18$pc at the distance of M87 (assumed to be at 15Mpc).
Since the absolute distances and the position angles of
all the positions are provided, we do not require $x_{in}$ and $x_{out}$
for each of these regions separately-- fixing a mass of the black hole
provides the coordinates $x$ and $\Psi$ for each of these positions.
For simplicity, while comparing results, we integrate the zone
confined by the spiral coordinates surrounding a circular region.
The successive (solid) circles are drawn with diameters of $0.26$ arc sec,
$0.78$ arc sec and $1.3$ arc sec respectively. The regions I-VI
are drawn at $\delta \Psi=60^{\deg}$ intervals and the regions
VII-XVIII are drawn at $\delta \Psi=30^{\deg}$ intervals.

In Fig. 1, we also present schematically the spiral shock waves
(thick curves) marked by $\Psi_{s1}$ and $\Psi_{s2}$
and the sonic surfaces marked by $\Psi_{c1}$ and $\Psi_{c2}$.
The direction of rotation is shown by an arrow. The line profiles
calculated in regions marked from I to XVIII are presented
in \S 4 as predictions of the model.

Observational results have been obtained by digitizing the line profiles
from Figs. 1b and 2 of H94 and converting the observed
wavelengths into wavelengths in the rest frame of the disk.
First we fit the profiles of Position 4
(nucleus) and check if the same set of parameters consistently reproduces
the profiles in Positions 1, 2, 5 and 6. Though different lines show
different behaviors in the observations, a general trend is found to
emerge which is the following: The line intensity from Position 4 is roughly
$2$ to $2.5$ times stronger than that from Positions 5 and 6
and about $4$ to $6$ times stronger than that from Positions 1 and 2.
The red shifted lines in Positions 5 and 6
are stronger than the blue-shifted components in some cases
(e.g. [NII] at $\lambda 6584$\AA  \ and [SII] at $\lambda 6731$\AA)
but weaker at other cases (e.g., [O III] lines at $\lambda\lambda
4959$\AA\ and $5007$\AA). Since our model is for general line behavior,
clearly, we cannot, for the same spiral structure, explain
these unequal asymmetries in different lines without making further assumptions
with respect to the inhomogeneities in the compositions,
the variation of the emissivity exponent $q$ and other physical
processes such as photoionization, which are not included here.
(In our model, intensity asymmetries can be different
in different position angle of the shock.)
We thus chose to ignore modeling the exact asymmetries
in different line intensities, and concentrated on the
basic features, such as the Doppler shift and the average behavior
of the intensity ratio from one position to the other.

While fitting the line profile from the nuclear region
(Position 4), we are faced with the
dilemma that whereas some lines clearly show splitting in the central
region (e.g., [NII] at $\lambda 6584$\AA \ and $H_\beta$ at $\lambda 4861$\AA)
others do not (e.g., [OIII] at $\lambda 5007$\AA). The latter
line widths however show indications of the split, as though they are
superpositions of the splitted lines with possibly some additional components
along the line of sight to smear out the valley.
In our model, we do see the splitting in the nuclear region in some
phases of the spirals. It is possible that different lines are
excited at different regions of the spiral and therefore `see' different
phases.

Regarding the sensitivity of the emitted lines on the
fitting parameters, we wish to mention that the emitted line profiles
are found to be very sensitive to the relative phase of the spiral
shocks with respect to the line of sight. Thus the spirals were
first rotated around the axis (for each set of above
mentioned parameters) and the results were found to be best
fitted when the shocks were placed as the thick curves in
Fig. 1 in relation to the observed positions. (The rotation by an angle
$\phi_s$ is done with respect to a fiducial direction
defined by $\phi^{\prime}=\Psi=0$ at $x=1$.)
Thus, we find that the results from Positions 5 and 6
may actually correspond to emissions from the hotter {\it post-shock}
regions. As the flow crosses the shock at $\Psi_{s1}$, it becomes
subsonic and hot, which produces the emissions in Position 6.
The flow subsequently becomes supersonic at the sonic surface at
$\Psi_{c1}$ and passes through the shock at $\Psi_{s2}$ producing the
emission in Position 5. This becomes supersonic again at $\Psi_{c2}$
and the cycle continues. Thus we believe that the spiral arms which are
observed in the $H_\alpha$+[NII] image (F94) actually represent post-shock
dissipative regions.

\section{Analytical Results and Comparison of Line Profiles}

A two-armed spiral shock solution in a self-similar disk forms within a limited
region of the parameter space spanned by the spiral angle $\beta$ and the
velocity coefficient $q_2$ at the sonic surface. The velocity coefficients
($q_i$s) as defined in Eqn. 1 remain constant on a given spiral curve but
vary with $\Psi$ and therefore, for a given $x$, on the azimuthal angle
$\phi^\prime$. In Figure 2, we present the variation of $q_{1-3}$ as functions
of the spiral coordinate $\Psi=\phi+$tan$\,\beta\ $log$(x)$, for $\beta \sim
9^o$ and $q_{2c}=0.54$. To obtain the actual velocity (in units of the
velocity of light) at a given point on the disk, one simply has to divide
these numbers by $x^{1/2}$, where $x$ is in units of $GM/c^2$, half of the
Schwarzschild radius. The curves are drawn  from the post-shock region
(left end) to the pre-shock (right end) in between two shock waves, as
the flow passes through the sonic surface ${\Psi=0}$. The sound speed
coefficient (long-dashed) increases from $0.53$ to $0.634$ at the shock,
a jump of about $20$\% which implies a jump of about $40$\% of the
temperature at the shock front. The azimuthal velocity coefficient
(short-dashed) jumps from $0.92$ to $0.27$ at the
shock front, indicating a strong dissipation of angular momentum
each time the flow passes the shock front (Paper 1). Notice that
$q_2 <1$ always, i.e., the flow is entirely sub-Keplerian. The
radial velocity coefficient (solid) goes down from $0.46$ to $0.36$
at the shock front while becoming negative in some region. Thus the
orbit is elliptical. The Mach number (defined with respect to the
velocity component normal to the shock) jumps from $1.86$ in the pre-shock
flow to $0.51$ in the post-shock flow, which signifies a jump by a factor
of $3.67$, i.e, the shock is relatively strong. Our procedure
finds the polytropic index $n$ self-consistently which in this case
happens to be $2.02$ and $\gamma \sim 1.5$. The angle between the sonic surface
and the preceding shock is $ \sim 120^{\deg}$, which
corresponds to the locations $\Psi_{c1,c2}$ drawn in Fig. 1.

In the following discussion, we shall choose this solution.
Before we present the comparison, we point out a simple but crucial
result: the azimuthal velocity varies from $27$\%
to about $92$\% of the Keplerian value with an average of roughly
$60$\%. As a result, if the effects of radial motion
on the line profiles are ignored, we should expect that the
required mass of the black hole to be higher by a factor of $\sim 0.60^{-2}
\sim 3$ to explain the same frequency shift. (This argument does not depend
upon the inclination angle or emissivity or the degree of thermal broadening.)
As we show below, in reality, we require the central mass to be around
$4 \times 10^9 M_\odot$, which is roughly twice
as high. This is also an indication that the radial motions play an important
role in shaping the line profiles.

Figure 3 shows the line profiles as obtained
from the theory (solid) and the observations (dashed) made from various
positions of the disk. The mass of the black hole is chosen to be $4 \times
10^9 M_\odot$ and the other parameters are:
$i=42^{\deg}$, $v_{th}=0.0005$ and $q=2.9$.
The dot-dashed curve is the Position 4 observation of
[NII] at $\lambda 6584$\AA\ (with $H_\alpha$ blended towards blue
widening the base considerably) and the short-dashed curves
are the Positions 4, 5, 6 observations of
[O III] line at $\lambda 5007$\AA. The observations from Positions 1 and 2
roughly match (eye estimation was made since no
detail line profiles were presented in F94 and H94)
the solid curves drawn for the corresponding positions.
We scaled vertically the observed [O III] line at Position 4 so that
it may match the intensity of the theoretical profile. The
scaling factor is then reduced by $35$\% to obtain the factor
for the observed lines at Positions 5 and 6 which are plotted
as well. We also scale separately the Position 4 observation of
[N II] (dot-dashed curve) so as to superpose this plot.
The [N II] lines at Positions 5 and 6 were heavily blended with
$H_\alpha$ lines and were not plotted. But it was noted that
these lines are red-shifted similarly as the theoretical solid
curves at these locations.

The general agreement of the observed line profiles and the theoretical curves
show that the mass of the black hole is possibly much higher than
the previously computed values. One important point to note is that
the line width of the wing for the Position 4 observation can be matched with
the theoretical profile only if the inner edge of the spiral shock
be chosen to be at around $x_{in}=2 \times 10^4$ (i.e., ten thousand
Schwarzschild radii). We find that for $x_{in}$ less than this value
the width of the lines at the wing becomes very broad.
Since temperature $T \propto x^{-1}$ in our model, it increases by
at least a factor of $10$ along curve of constant $\Psi$, as one moves in
from $1$ arc sec to $0.1$ arc sec distance from the center.
Thus, either the shocks do not extend below this radius, or, if they do,
they become so hot that line emission properties change dramatically.

We see above that the constraints from the wing width
indicate $M(R<x_{in}) \sim 4 \times 10^9 M_\odot$ which is a more
stringent limit on the central mass than previously presented.
It could be easily shown that the results do not match with observations
if the mass of the black hole is chosen very much differently. If the black
hole
is chosen to be lighter, the line shifts are found to be
much less than the observed values. Similarly, with a higher mass, the line
shifts are found to be higher. Based on the nature of the fit, we believe
that the mass of the black hole and the inclination angle could be within the
range $M=4.0 \pm 0.2 M_\odot$ and $i=42\pm 2^{\deg}$ respectively. Thus the
mass
estimate from our calculation is accurate to a few percent. This is to be
contrasted with the mass estimate of H94, where $\pm 30$\% error bar
in the mass is obtained by forcing the flow to have Keplerian velocity.

In Figures 4(a-b) we present the predictions of the line profile from
the theoretical model with the same parameters. In Fig. 4a, lines
from regions marked I-VI and in Fig. 4b, lines from the regions
marked VII-XVIII are presented. Because of the non-axisymmetric nature of the
flow, the line intensities and the frequency shifts do not vary
monotonically. Future observations from these regions with
similar aperture can be used to verify such a flow behavior in the disk.

In anticipation of observations with $0.09$ arc sec aperture, we present in
Figure 5 the predicted behavior of the line profile closer to the
center. We chose regions (a-f) with diameter $0.09$ arc sec, centered
at $0.135$ arc sec, and with position angles $81^{\deg}$, $144^{\deg}$,
$201^{\deg}$, $261^{\deg}$, $321^{\deg}$ and $21^{\deg}$ respectively.
These positions are radially below positions I to VI of Fig. 1.

\section{Concluding Remarks}

We have modeled the ionized disk in M87 assuming that
the disk has two stationary spiral shock waves. The line
profiles computed from this model generally agree with those
recently observed by Hubble Space Telescope (F94, H94)
provided the mass of the central black hole is about $4 \times 10^9 M_\odot$.
Mass evaluated from our model is higher compared to the previously
estimated values. This is primarily due to the fact that the azimuthal
velocities in disks with shock waves are sub-Keplerian
and therefore one requires a higher central mass for the same frequency shift.
Our analysis, however, agrees with the estimated inclination
angle $i \sim 42^{\deg}$. Another significant observation is that the
spiral arms which are observed in $H_\alpha$+ [NII] images (H94, F94)
may actually correspond the post-shock emission regions
in the theoretical model where the observations 5 and 6 (H94, F94) were made.
We also make prediction of the variation in line shapes from different
positions in the disk, and thus our model can be tested by future observations.
If this sub-Keplerian disk model is correct, it may generally indicate that
so far, the masses of the galactic nuclei are seriously under estimated. This
may have other ramifications, such as the constraint on the
the dark matter candidates in galaxies and in the universe in general.
Also, the existence of such a super-massive black hole at the galactic
nucleus may indicate that since the time of catastrophic
formation of the black hole during the proto-galactic phase,
the central mass may have significantly increased through accretion
of gas and tidal capture of stars.

Though we have advanced the model of line emission from shock heating,
it  is possible that some degree of photoionization may not be ruled
out. One way to distinguish different emitting region is to study
the intensity ratios of different lines (Veilleux \& Osterbrock, 1987).
{}From the HST data (H94), this ratio is found to vary from position to
position. Generally,  the ratios are:
$([N\ II] \lambda 6583)/(H_\alpha\ \lambda 6563) \sim 3.0$,
$([O\ III] \lambda 5007)/(H_\beta\ \lambda 4861) \sim 1.5\ -\ 2.2$,
$([S\ II] \lambda 6716 + \lambda 6731 )/(H_\alpha\ \lambda 6563) \sim 0.9\
-\ 1.0$, $([O\ I] \lambda 6300)/(H_\alpha\ \lambda 6563) \sim 0.5$.
These ratios definitely lie in the AGN region  and not in H II region
(see Figs. 4-6 of Veilleux \& Osterbrock, 1987).
Whereas the plot of $OIII/H_\beta$ vs.
$SII/H_\alpha$ falls close to the shock heating curve of Shull \& McKee (1979),
other plots fall on the right of it and could possibly be fitted with
photoionization models (e.g., Ferland \& Netzer, 1983; Stasinska 1984) as well.
Infact, the situation is similar to the case of LINERs where it is difficult to
distinguish the origin from the ratio alone. One the one hand, Keplerian disks
would be very cold at distances of $10^5$ Schwarzschild radius to emit the
observed lines. On the other hand, if the flow is really transonic as in
our model, one draws an unavoidable conclusion that the sound velocity
is several hundred Km s$^{-1}$ at this distance, and therefore the tempereture
in equatorial plane is $\sim 10^7$K! Thus, the disk is probably a mixture
of ionized and neutral components with only a fraction of matter
participating in the process of shock formation.

We have already mentioned that the vertical equilibrium model
may breakdown in the post-shock region where some vertical
motion is expected after the flow thickens at the shock.
In the present case, the shock thickness ($\propto q_3$)
is expected to be higher by only 20\% from that of the immediate pre-shock
flow. In reality, in presence of viscosity and the resulting
smearing effects this velocity need not be high. Furthermore, the ratio
$h/x = q_3 =0.64 < 1$ in the post-shock region so that our treatment
is probably justified. A corollary of the thickening of the disk
at the shock is the shadowing effect which is expected to obscure the
pre-shock disk. However, we have ignored this effect for two reasons:
(a) the shock is almost straight
and is located only a few degrees away from the projected line of sight
on the disk plane and (b) the positions 5 and 6 fall
in the post-shock region, while the positions 1 and 2 are far away from it
and therefore free from shadowing effects. If the shadowing effect is
important (namely, if the thickening is not gradual but abrupt), the resulting
line profile, emitted from regions IV \& XIII would have intensity a few
($\leq 5$) percent smaller than what is predicted in Figs. 4(a-b).
We notice from the $H_\alpha$+[NII] image (F94) of the spiral arms extended
beyond $1$ arc sec that the spiral waves definitely are not self-similar.
They suddenly bend by a higher angle. However, within about $1$ arc sec,
they are roughly straight and thus may actually correspond to the post-shock
regions of the $\beta \sim 9^{\deg}$ shocks which we model here.
We also conclude that the properties of the shock and/or the emission
property of the disk must be different very close to the nucleus, since
extension of the shocks below $10^4$ Schwarzschild radius causes the line
widths to be unduly higher and the theoretical results
deviate from observations considerably.

The author is grateful to H. Ford and A. Kochhar for allowing him
to have copies of their papers (F94 and H94) before publication.
He thanks P. Schneider for carefully reading the manuscript and F. Mayer
and M. Camenzind for helpful comments.
The author is also thankful to the Indian National Science Academy
and Deutsche Forschungsgemeinschaft which made his visit to the Max Planck
Institute possible.

\vfil\eject

\centerline {FIGURE CAPTIONS}
\noindent Fig.1: {Positions (1, 2, 4, 5, 6) of the observed region of the
disk, locations of the shock waves ($\Psi_{s1,s2}$) and
sonic surfaces ($\Psi_{c1,c2}$). Also shown are regions I-XVIII.
Predicted line profiles from these regions are presented in Figs. 4(a-b)}

\noindent Fig. 2: {Variations of the radial (solid), azimuthal (short-dashed)
and sound (long-dashed) velocity coefficients $q_i$ as
functions of the spiral coordinate $\Psi$ from the post-shock region (left
end) to the pre-shock region (right end). $\Psi=0$ represents the sonic
surface.

\noindent Fig. 3: {A comparison of the model (solid curves) and observed
(dot-dashed for [NII] $\lambda 6584$\AA\ and short-dashed for [O III]
$\lambda 5007$\AA)
line profiles plotted against frequency. The model parameters are:
$M=4 \times 10^9 M_\odot$, $\beta = 9.1^{\deg}$, $q_{2c}= 0.54$, $q=2.9$, $i=
42.0^{\deg}$, $v_{th}=0.0005$. Model curves are marked with Position numbers
on the disk (see Fig. 1)}

\noindent Fig. 4(a-b): Predicted line profiles from the regions
(a) I-VI and (b) VII-XVIII. See Fig. 1 for definitions of these regions.

\noindent Fig. 5: Predicted line profiles of regions
centered at $0.135$ arc sec and $0.09$ arc sec diameter. The position
angles of the profiles (a-f) are $81^{\deg}$, $144^{\deg}$,
$201^{\deg}$, $261^{\deg}$, $321^{\deg}$ and $21^{\deg}$ respectively.


\begin{references}

\reference Chakrabarti, S.K. 1990a, ApJ  362, 406   (Paper 1)
\reference Chakrabarti, S.K. 1990b, Theory of Transonic Astrophysical Flows,
(Singapore: World Scientific)
\reference Chakrabarti, S.K. \& Matsuda, T. 1992, ApJ, 390, 639
\reference Chakrabarti, S.K., \& Wiita, P.J. 1993, ApJ, 411, 602
\reference Chakrabarti, S.K., \& Wiita, P.J. 1994 ApJ, (in press)
\reference Chen, K., Halpern, J.P., \& Fillipenko, A. 1989, ApJ,
339, 742 (CHF)
\reference Ferland, G.J. \& Netzer, H. 1983, ApJ, 264, 105
\reference Ford, H.C., Harms, R.J., Tsvetanov, Z.I., Hartig, G.F., Dressel,
L.L., Kriss, G.A., Davidsen, A.F., Bohlin, R., \& Margon, B. 1994, ApJ Letters,
(in press)
\reference  Harms, R.J., Ford, H.C., Tsvetanov, Z.I., Hartig, G.F., Dressel,
L.L., Bohlin, R., Kriss, G.A., Davidsen, A.F., Margon, B. \& Kochhar, A.
1994, ApJ Letters, (in press)
\reference Matsuda, T., Inoue, M., Sawada, K., Shima, E., and Wakamatsu, K.
1987, MNRAS, 229, 295
\reference Shull, J.M. \& McKee, C.J. 1979, ApJ, 227, 131
\reference Smak, J. 1981,  Acta Astron.  31, 25
\reference Spruit, H.C. 1987, A \& A, 184, 173
\reference Stasinska, G. 1984, A \& A, 135, 341
\reference Tylenda, R. 1981, Acta Astron.  31, 127
\reference Veilleux, S. \& Osterbrock, D.E. 1987, ApJ Supp. Ser., 63, 295

\end{references}
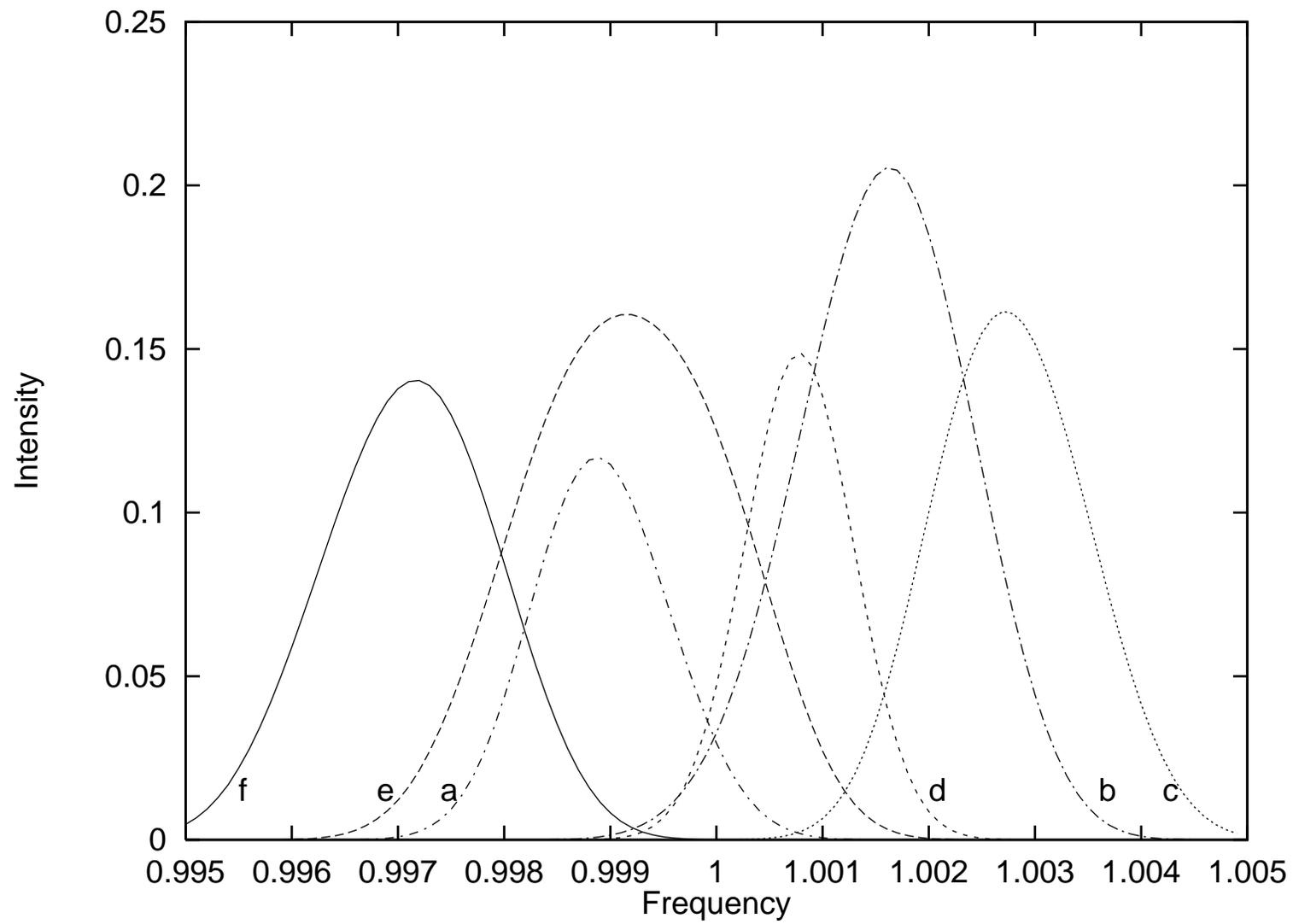
\end{document}